# Correlation between two non-linear events in a complex dusty plasma system

Rinku Mishra[1,2], S. Adhikari[2], Rupak Mukherjee[3], M. Dey[2]

[1]*Physics Department, Gauhati University, Guwahati 781014, India*

[2]*Centre of Plasma Physics, Institute for Plasma Research, Nazirakhat, Sonapur 782402, Assam, India*

[3]*Institute for Plasma Research, HBNI, Bhat, Gandhinagar 382428, Gujarat, India*

**Abstract**

A phenomenological model using fluid theory is developed to show that the decay of two seemingly independent nonlinear structures namely the dust void and dust soliton strongly depends on the plasma ionization parameter in an unmagnetized complex plasma system. Numerical solution of model equations has shown that the evolution of dust voids and their subsequent decay in a time frame is intimately related with ionization parameter. A similar result also holds good in case of dust solitons where stability of soliton is found to depend critically upon ionization parameter. Most importantly, it is observed that time of the collapse of a dust soliton precedes the onset time of a dust void decay and therefore soliton decay acts as a precursor for void decay to occur in a given dusty plasma system.

## I. INTRODUCTION

From very early days in science, researchers have always tried to dig deeper into fascinating natural phenomena. Such observations are mostly non-linear in nature and have a profound effect on its surrounding as well as on the system. Dusty plasma is one such interesting area where various non-linear events have been observed experimentally. The dust particles present in the plasma system get negatively charged primarily due to the highly mobile electrons present in the plasma. The surprising linear, non-linear and chaotic dynamics and exotic structure formations in the dusty plasma medium has attracted various researchers into this field in the last two decades. Such processes in a dusty plasma include nonlinear dispersion relation, solitons, shock wave, formation of self-organized structure (dust void formation etc.) and many more.

The present paper aims to study two apparently different non – linear phenomena, dust void formation and soliton propagation in a dusty plasma medium and attempts to establish a correlation between them. The dust void formation and solitary wave structure in any complex plasma medium are well-explained phenomena. There are several theories and experiments in support of the formation of dust void[1–4] and propagation of solitary wave structures[5–8] in non-linear dusty plasma systems.

Dust void, a dust free region inside the dusty plasma, was first observed in an experiment performed by Praburam and Goree[2]. Such structures form in a uniform dust cloud when dust particle density is very large and supported by an increased rate of ionization [9,10]. The theory[1] suggests that with the increment of ionization a hot spot with positive space charge is generated in the central region. The generated electric field exerts two forces namely electrostatic and ion drag force on the dust particles. The point of balance of these two forces represent the boundary of the void. While the electron and ion density remain continuous, a new type of discontinuity arises due to the rapid change in the dust density profile showing the presence of a sharp boundary separating the dust void and the dusty plasma region. Since the percentage of ionization in the plasma medium plays a critical role in the formation of the void structure, the present study intends to investigate the dependence of void formation on ionization as a parameter. In a similar study, Vladimirov et al.[11] tried to examine the stability of steady-state void by considering the ion-neutral collisions as well as the ionization factor. The study concluded that the position of zero net force on dust particles represents the dust void boundary, which is considered as the basis of void theory in a dusty plasma. The time-dependent model for dust void was first developed by K. Avinash et al.[12] in 2003. The model confirms that the dust void arises due to the instability caused by the ion drag force and its size depends on the ionization value. The electric field was found to be very low in the central region and increases gradually while moving away from the center. It attains maxima at the boundary where it balances the ion drag force. There is a common practice to study the void formation in dusty plasmas using a steady-state system of equations[1,4,11]. However, the time-dependent approach reveals more insight about the physical process taking place in such system [3,12]. Using a time-dependent approach, the evolution of the dust density can be investigated including the effect of ion drag force, diffusion of dust, ion-neutral and dust-neutral collisions. Such studies focus more on the parameters that affect the stability of dust void and its dependency with non-linear solitary structure. Moreover, the force acting on the dust void structure can confirm the position of the void.

Solitons, on the other hand, manifest in a nonlinear medium as a finite amplitude wave that maintains its solitary stable structure because of balance between wave steepening and dispersion during propagation and collision. Solitons are widely studied in non-linear systems e.g., in hydrodynamics, plasma physics, condensed matter physics, and in optics. The conventional method to study the ion acoustic nonlinear wave phenomena is to derive the KdV equation using the Reductive perturbation techniques (RPT)[5]. In plasma physics, solitons are generally derived from KdV equation in the form of ion-acoustic and dust acoustic modes[13,14].

In this paper, we have undertaken an interesting study in which the primary focus is to develop an analytical relation to show the linkage between the stability of two nonlinear structure viz. a dust void and a soliton in an unmagnetized dusty plasma system. The paper is arranged as follows. The basic equations and modeling are presented in Sec. II. Sec. III contains the numerical parameters that have been used in this study. Results and discussions are provided in section IV. Finally, Sec. V contains the brief conclusions of the paper.

## II. BASIC EQUATIONS AND MODELING

The model consists of time-dependent unmagnetized dusty plasma system based on a set of fluid equations. The micron-sized dust grains immersed into the plasma behaves as a collisional charged fluid, where the electrons due to their negligible mass possess high mobility thereby forming a Boltzmann distribution around the dust particles. The time scale of dust motion is quite small as compared to the electrons and the ions. It is assumed that ion density can quickly adjust itself to its initial constant value within the slow time scale of dust motion. Therefore, the flow of ion density is considered to be a constant quantity in this model[12]. Although the assumption of constant ion density seems unusual, the present context demands a constant supply of ions to sustain the continuous flow in the void. The existing assumption ensures to fulfill the same.

When a neutral gas becomes ionized, the ionized fraction exhibits electromagnetic properties and behave as a plasma. The ionization process is mostly dominated by ion-neutral collision in an electron-ion plasma. However, in a dusty plasma medium, dust plays an important role and therefore incorporation of exact ionization in the fluid equation for a particular plasma system is more complex than it seems, especially in the present situation where we cannot put the ionization term explicitly in the ion momentum equation. The ionization affects the dust flow in the system involving more complexity in the void stability. Therefore, the role of ionization term in the governing equation is quite subtle in nature. The system of equations have been adapted from the time-dependent void model[12] with the addition of convective term in the system.

It is important to mention here that the phenomena we are trying to resolve is independent of any spatial dimension. The previous studies have shown that dust void as well as solitons can be studied with higher degree of accuracy using 1D modelling[1,5,11]. Such 1D study also minimizes the complicacy arising from spatially dependent plasma parameters.

The set of fluid equations containing the continuity, and momentum equations for dust particles are as follows:

$$\frac{\partial n_d}{\partial t} = -\frac{\partial (n_d v_d)}{\partial x} + D\frac{\partial^2 n_d}{\partial x^2} \tag{1}$$

$$m_d \frac{\partial v_d}{\partial t} - v_d \frac{\partial v_d}{\partial x} = Ze\frac{\partial \varphi}{\partial x} + F_d - \nu_{dn} m_d v_d - \frac{T_d}{n_d}\frac{\partial n_d}{\partial x} \tag{2}$$

$$\frac{T_e}{n_e}\frac{\partial n_e}{\partial x} = -n_e E \tag{3}$$

$$u_i = \frac{eE}{m_i \nu_{in}} \tag{4}$$

The Poisson's equation for the system is given as

$$\frac{\partial^2 \varphi}{\partial x^2} = -4\pi e(n_i - n_e - Zn_d) \tag{5}$$

$n_d, n_i,$ and $n_e$ are the dust, ion and electron density, the second term of Equation (1) in the right side represents the diffusion of dust particles. Z is the amount of charge on the dust particles. $F_d$ is the approximated non-linear ion drag force[12] given by $F_d = m_d \nu_{di} v_{thi} u/(b + u^3)$, $\nu_{di}$ is the ion dust collision frequency and $b$ ( $= 1.6$ ) is a constant. Third and fourth term of Equation (2) represents the friction of dust and neutral particles, and dust pressure. $\nu_{dn}$ is the dust neutral collision frequency. The normalized set of the above fluid equations are given as

$$\frac{\partial N_d}{\partial T} = -\frac{\partial (N_d V_d)}{\partial X} + D_0 \frac{\partial^2 N_d}{\partial X^2} \tag{6}$$

$$\frac{\partial V_d}{\partial T} - V_d \frac{\partial V_d}{\partial X} = \frac{\partial \psi}{\partial X} + \frac{au}{b + u^3} - \alpha_0 V_d - \frac{\delta}{N_d}\frac{\partial N_d}{\partial X} \tag{7}$$

$$\frac{\partial N_e}{\partial X} = \tau_i^{-1} N_e \frac{\partial \psi}{\partial X} \tag{8}$$

$$\frac{\partial^2 \psi}{\partial X^2} = -(1 - N_e - N_d) \tag{9}$$

$$u = -\mu \frac{\partial \psi}{\partial X} \tag{10}$$

### A. Normalized parameters

Dimensionless quantities used in the system are as follows: $X = (\tau_i x)/\lambda_i$, $T = \omega_{pd} t$, $\psi = e\varphi/T_e$, $N_d = Zn_d/n_{i0}$, $N_e = n_e/n_{i0}$, $\alpha_0 = \nu_{dn}/\omega_{pd}$, $\tau_{i,d} = T_{i,d}/T_e$, $\delta = \tau_d/Z$, $\mu = \omega_{pi}/\nu_{in}\tau_i$, $D_0 = D\tau_i^2 \omega_{pd}/\lambda_{di}^2$, $a = (m_d \nu_{di})/(m_i Z \nu_{in})$.

### B. Dust void Solution

The geometrical distribution of unmagnetized plasma is considered symmetric with respect to a given origin in a one dimensional system. The self consistently time evolved set of fluid equations (6-10) has been numerically solved to obtain the evolution of dust void in space and time as shown in figure 2 (a). Initial values used in the problem are $N_d = 0.001$, $N_e = 0.999$, $N_i = 1$, $E = 0.001$, $V_d = 0$. Numerical solution of model equations has revealed that the stability of dust void depends critically on the strength of the plasma ionization parameter ($\mu$). The role of plasma ionization parameter ($\mu$) in defining the stability of dust void and soliton structure is presented in Sec. III, with more details.

### C. Solitary structure

Using the above set of fluid Equations (6-10), the KdV equations have been obtained by using the reductive perturbation technique. The stretched parameters[15] used in the system are

$$\xi = \epsilon^{\frac{1}{2}}(X - \lambda T) \quad \text{and} \quad \eta = \epsilon^{3/2} T$$

$\epsilon$ is the small dimensionless expansion parameter and $\lambda$ is the soliton's velocity.

Expansion of all the independent variables in terms of small perturbation $\epsilon$ are as follows:

$$N_j = N_{j0} + \epsilon N_{j1} + \epsilon^2 N_{j2} + \ldots \ldots \quad j = e, d \tag{11}$$

$$\psi = \psi_0 + \epsilon \psi_1 + \epsilon^2 \psi_2 + \ldots \ldots \tag{12}$$

$$V_d = V_{d0} + \epsilon V_{d1} + \epsilon^2 V_{d2} + \ldots \ldots \tag{13}$$

$$u = u_0 + \epsilon u_1 + \epsilon^2 u_2 + \ldots \ldots \tag{14}$$

Substituting Equations (11-14) into above Equations (6-10) and collecting the lowest order coefficients of $\epsilon$ we have

$$N_{d1} = -\frac{N_{d0}}{V_{d0} - \lambda}V_{d1} \quad \text{and} \quad V_{d1} = \left(\frac{(N_{d0}b - a\mu N_{d0})(V_{d0} - \lambda)}{b\left[(N_{d0}N_{d0} - N_{d0}\lambda)(V_{d0} - \lambda) - \delta N_{d0}\right]}\right)\psi_1$$

$$\text{Therefore,} \quad N_{d1} = -\frac{N_{d0}}{V_{d0} - \lambda}\left(\frac{(N_{d0}b - a\mu N_{d0})(V_{d0} - \lambda)}{b\left[(V_{d0}N_{d0} - N_{d0}\lambda)(V_{d0} - \lambda) - \delta N_{d0}\right]}\right)\psi_1$$

(15)

$$\text{and} \quad N_{e1} + N_{d1} = 0$$

Next, equating the higher order coefficients of $\epsilon$, we have

$$-\frac{\partial N_{d2}}{\partial \xi} = -\frac{N_{d0}}{V_{d0} - \lambda}\frac{\partial V_{d2}}{\partial \xi} - \frac{N_{d1}}{V_{d0} - \lambda}\frac{\partial V_{d1}}{\partial \xi} - \frac{V_{d1}}{V_{d0} - \lambda}\frac{\partial N_{d1}}{\partial \xi} - \frac{1}{V_{d0} - \lambda}\frac{\partial N_{d1}}{\partial \eta} \tag{16}$$

$$\frac{\partial V_{d2}}{\partial \xi} = \frac{\lambda}{V_{d0}N_{d0} - \lambda N_{d0}}N_{d1}\frac{\partial V_{d1}}{\partial \xi} - \frac{V_{d0}}{V_{d0}N_{d0} - \lambda N_{d0}}N_{d1}\frac{\partial V_{d1}}{\partial \xi} - \frac{N_{d0}}{V_{d0}N_{d0} - \lambda N_{d0}}\frac{\partial V_{d1}}{\partial \eta} + \frac{\left(N_{d0} - \frac{a\mu N_{d0}}{b}\right)}{V_{d0}N_{d0} - \lambda N_{d0}}\frac{\partial \psi_2}{\partial \xi} + \frac{\left(1 - \frac{a\mu}{b}\right)}{V_{d0}N_{d0} - \lambda N_{d0}}N_{d1}\frac{\partial \psi_1}{\partial \xi} - \frac{\delta}{V_{d0}N_{d0} - \lambda N_{d0}}\frac{\partial N_{d2}}{\partial \xi} - \frac{V_{d1}N_{d0}}{V_{d0}N_{d0} - \lambda N_{d0}}\frac{\partial V_{d1}}{\partial \xi}$$

(17)

$$\frac{\partial N_{e2}}{\partial \xi} = \tau_i^{-1}N_{e0}\frac{\partial \psi_2}{\partial \xi} + \tau_i^{-2}N_{e0}\psi_1\frac{\partial \psi_1}{\partial \xi} \tag{18}$$

$$\frac{1}{\lambda^2}\frac{\partial^2 \psi_1}{\partial \xi^2} = N_{e2} + N_{d2} \tag{19}$$

Using Equations (16-19), the KdV equation obtained can be written as

$$\frac{\partial \psi_1}{\partial \eta} + P\psi_1\frac{\partial \psi_1}{\partial \xi} + Q\frac{\partial^3 \psi_1}{\partial \xi^3} = 0 \tag{20}$$

where, P and Q are given as $\quad P = N/M \quad \text{and} \quad Q = -1/M$

with,

$$M = \frac{-(V_{d0}N_{d0} - \lambda N_{d0})B + N_{d0}^2 A}{\left[(V_{d0}N_{d0} - \lambda N_{d0})(V_{d0} - \lambda)\right] - \delta N_{d0}}$$

$$N = \frac{-2(V_{d0}N_{d0} - \lambda N_{d0})AB - N_{d0}\lambda AB - N_{d0}\left(1 - \frac{a\mu}{b}\right)B + N_{d0}^2 A^2 + V_{d0}N_{d0}AB}{\left[(V_{d0}N_{d0} - \lambda N_{d0})(V_{d0} - \lambda)\right] - \delta N_{d0}} + \frac{N_{e0}}{\tau_i^2}$$

$$A = \frac{((N_{d0}b - a\mu N_{d0})(V_{d0} - \lambda))}{b[(V_{d0}N_{d0} - N_{d0}\lambda)(V_{d0} - \lambda) - \delta N_{d0}]}$$

$$\text{and} \quad B = -\frac{N_{d0}A}{V_{d0} - \lambda}.$$

In terms of $N_{d1}$ the KdV Equation (20) is reframed as,

$$\frac{\partial N_{d1}}{\partial \eta} + \beta N_{d1}\frac{\partial N_{d1}}{\partial \xi} + \alpha \frac{\partial^3 N_{d1}}{\partial \xi^3} = 0 \tag{21}$$

where, $\beta = \frac{P}{B}$ and $\alpha = Q$.

The analytical solution of equation (21) is as follows

$$N_{d1} = \frac{3\lambda}{\beta}\text{sech}^2\left(\frac{1}{2}\sqrt{\frac{\lambda}{\alpha}}\right)X \tag{22}$$

where $X = \xi - \lambda T$.

### III. NUMERICAL METHOD AND PARAMETER DETAILS

The second order finite difference discretization method has been used to simulate the dynamics of dust void governed by equations (6-10). The time-dependent problem has been taken care of by using the central difference in space and the forward difference in time. On the other hand, the KdV equation (21) for the solitary wave has been solved by using the Fourier Spectral method[16] due to its better stability and higher accuracy. The complete numerical process has been done using Matlab.

The hydrodynamic fluxes are usually discretized using explicit methods. The Courant–Friedrichs–Lewy (CFL) number of the hydrodynamic system limits the stability of such scheme. Therefore, it is crucial to choose the numerical parameters cautiously in order to satisfy the condition. The numerical parameters considered in the fluid simulation are as follows:

System length ($L_x$) = 8 with $N_x$ = 100 and time stepping ($\Delta t$) = $10^{-4}$. The CFL number comes around 0.000462 << 1.

The parameters used in the calculation are $m_d = 1.414 \times 10^{-18} kg$, $T_d = 0.02 eV$, $D_0 = 0.01$, $\delta = 0.001$, $m_i = 6.6 \times 10^{-26} kg$, $\nu_{dn} \sim 150 s^{-1}$, $\omega_{pd} \sim 700 s^{-1}$, $\lambda = 0.05 m$, $n_{e0} = 5 \times 10^{16} m^{-3}$, $n_{i0} = 2.5 \times 10^{16} m^{-3}$, $Z = 200$, $r \sim 10^{-6} m$, $n_{d0} = 1.25 \times 10^{14} m^{-3}$, $\tau_i = 0.125$.

## IV. RESULTS AND DISCUSSIONS

The model attempts to study the evolution of two seemingly unrelated nonlinear structures viz. dust void and soliton in a homogeneous unmagnetized dusty plasma system by using fluid theory in the same frame. Although, the physical mechanism governing the stability of dust voids and solitons has no commonality, however, in the present model we have seen that ionization parameter ($\mu$) is mainly responsible for the stability of dust void. For the lower value of $\mu$, void structure is short lived and for the higher value of $\mu$, stable void structure persists for a longer time. This obvious observation stems from the fact that higher ionization rate is essential inside the void for the creation of strong inward electrostatic force on a dust particle to counter the outward ion drag force at any point of time.

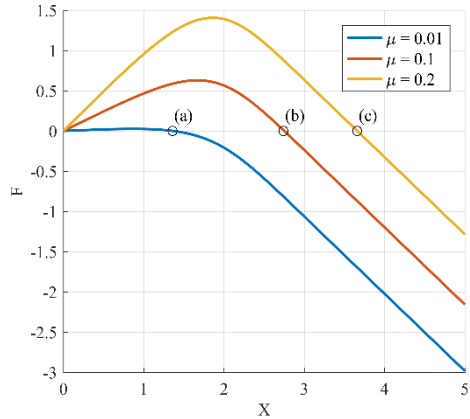

FIG. 1. The force balance (F) acting on the dust particle for different value of ionization ($\mu$). The point of zero force (a, b, c) represents the position of the dust void boundary for corresponding values of $\mu$.

Figure 1 shows the spatial profile of the force balance acting on a dust particle in the dust void system. The zero force represents the balance of the electrostatic and the ion drag force on a dust particle and this represents the boundary of the void. Points marked as *a, b,* and *c* represent dust void boundary for $\mu = 0.2$, $\mu = 0.1$, and $\mu = 0.01$ respectively. It is noted from the figure 1 that void size depends directly on ionization parameter $\mu$, and larger voids are formed for higher values of $\mu$. Moreover, void size can be determined from the figure 1 by observing the force balance points. Figure 2, (a) represents the dust density profile before breaking and (b) represents the dust density profile after breaking.

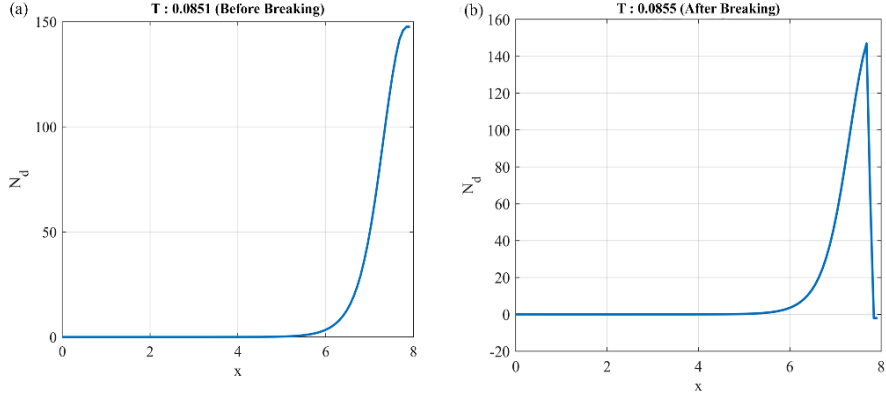

FIG. 2. The dust density profiles ($N_d$) for the ionization value, $\mu = 0.01$ (a) before breaking (T = 0.0851) and (b) after breaking (T = 0.0855).

Figures (3a), (4a) and (5a) show the breaking points of the stable void structure for different values of ionization parameter ($\mu$). For small $\mu$, the void structure remains stable for a shorter duration (T: 0.0631 for $\mu = 0.01$), but remains stable for a longer time of T : 0.855 when $\mu = 0.2$.

In order to have a comparative study with the stability of soliton structure in the same dusty plasma system, we have investigated the existence of small amplitude solitary wave structure by spectral solution of the KdV equation (21). It is seen that the ionization parameter affects the stable structure of solitary waves as well. The evolution of the solitary wave is shown in figures (3b), (4b) and (5b). From these waterfall plots, it is clear that for smaller values of ionization, the stability of the solitary structure is moderately short. However, for higher ionization value, the stability of the soliton increases.

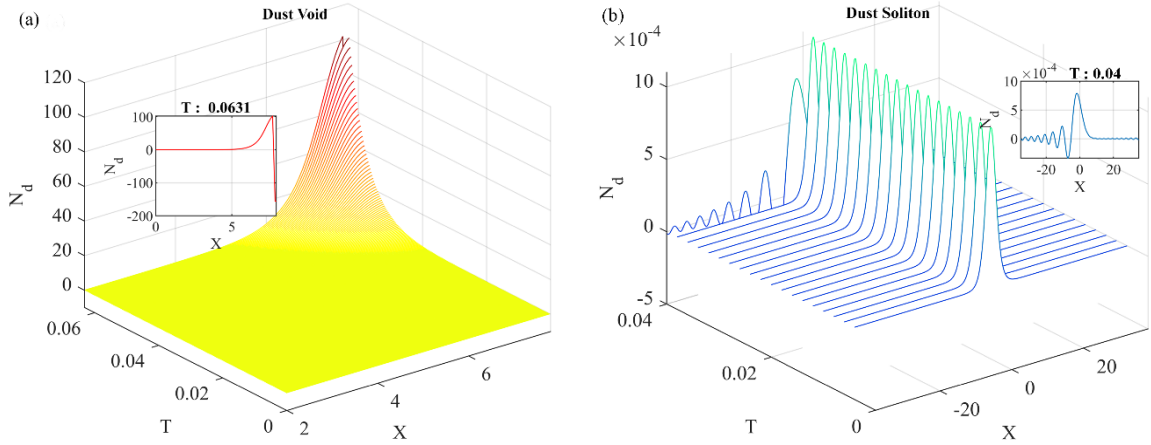

FIG. 3. (a) Evolution of dust density ($N_d$) with time (T) for the ionization value, $\mu = 0.01$. Inset figure represents the dust density profile at the breaking point ($T_b^v = 0.0631$) of dust void. (b) Solitary wave structure in the same dusty plasma system for the same ionization value ($\mu$). Inset figure represent the time ($T_b^s = 0.04$) and breaking pattern of solitary wave structure.

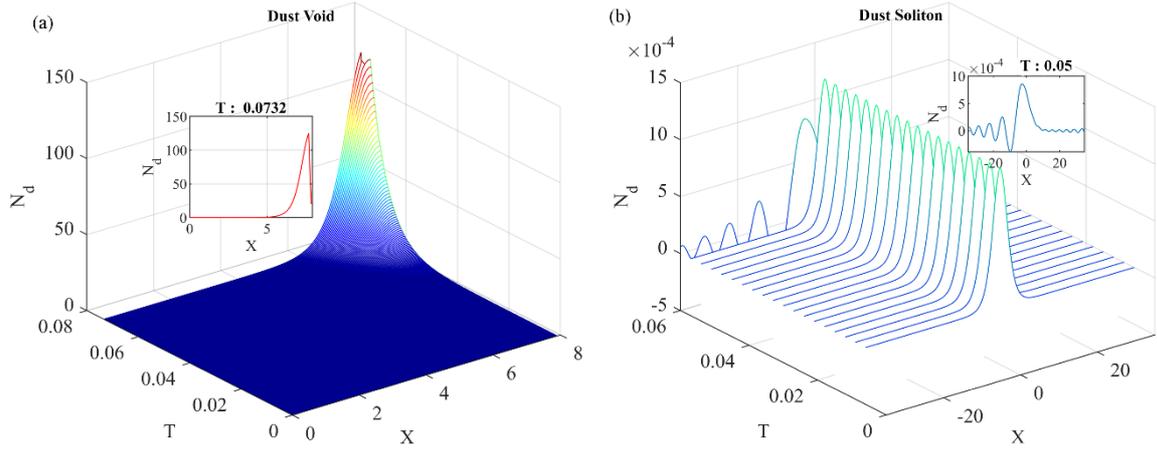

FIG. 4. (a) Evolution of dust density ($N_d$) with time ($T$) for the ionization value, $\mu = 0.1$. Inset figure represents the dust density profile at the breaking point ($T_b^v = 0.0732$) of dust void. (b) Solitary wave structure in the same dusty plasma system for the same ionization value ($\mu$). Inset figure represent the time ($T_b^s = 0.05$) and breaking pattern of solitary wave structure.

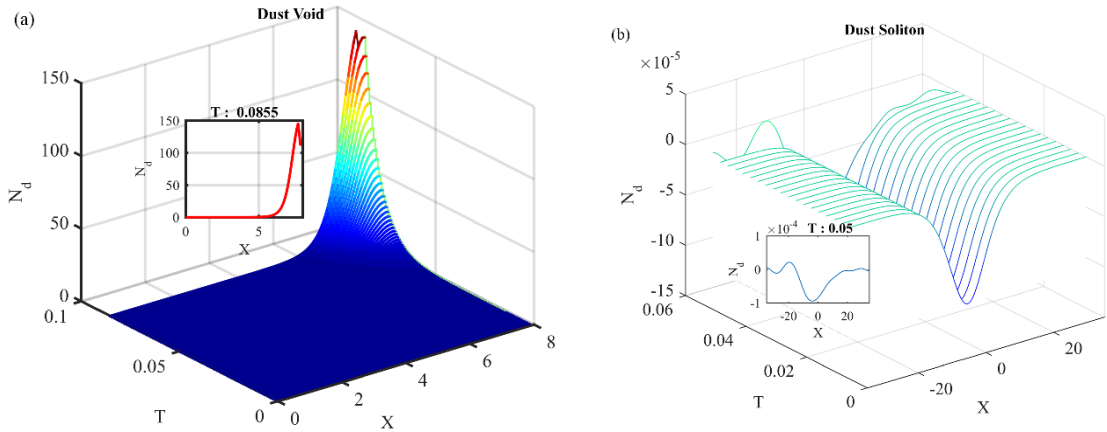

FIG. 5. (a) Evolution of dust density ($N_d$) with time ($T$) for the ionization value, $\mu = 0.2$. Inset figure represents the dust density profile at the breaking point ($T_b^v = 0.0852$) of dust void. (b) Solitary wave structure in the same dusty plasma system for the same ionization value ($\mu$). Inset figure represent the time ($T_b^s = 0.06$) and breaking pattern of solitary wave structure. The explanation of the negative dust density is provided with detail in the text below.

The inset figures in Figure (3b), (4b) and (5b) show the breaking pattern of the solitary wave structure for different ionization values ($\mu$). Irrespective of the ionization values, the patterns indicate that the solitary wave structures decay asymptotically and disappear thereafter. One important feature to observe here is the pattern of a solitary wave when $\mu = 0.2$. It can be seen from the Figure (6a) that the sign of non-linear coefficient ($\beta$) depends on the value of the ionization factor. For a higher value of $\mu$, the non-linear coefficient becomes negative and soliton becomes rarefactive (Figure (5b))[17].

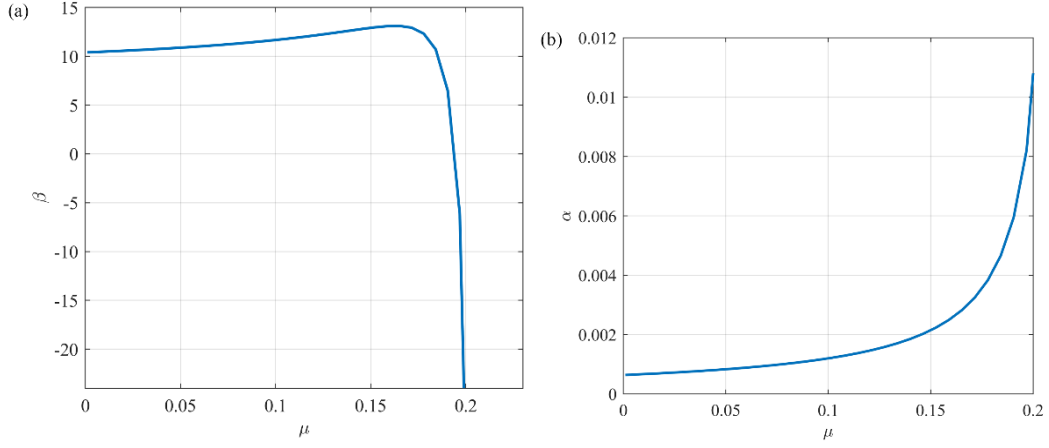

FIG. 6. The variation of (a) nonlinear coefficient ($\beta$) and (b) dispersion relation ($\alpha$) with ionization factor ($\mu$) in solitary wave structure.

Since solitons are formed due to balance between wave steepening (nonlinearity) and dispersion in the system, we have investigated the effect of ionization parameter on wave nonlinearity ($\beta$) and dispersion. Numerical analysis has shown that the ionization parameter $\mu$ has a negligible effect on wave dispersion ($\alpha$), however, wave nonlinearity gets drastically affected even for a small change in $\mu$ (Figure 6).

In order to establish a correlation between these two non-linear events, viz. dust void and dust soliton, it is important to observe both the phenomena simultaneously in the system. A close observation of figures (3-5) shows that the breaking of dust void and soliton structures occur with a phase difference. Both the phenomena are strongly affected by the ionization parameter $\mu$ and the breaking point of both the structure increase for higher values of $\mu$. It is also observed that the breaking of the solitary wave structure always takes place before the collapse of dust void structure. Hence, the breaking time of a soliton structure seems to act as a precursor for the decay of a dust void for a given $\mu$. Thus, we see that there exists a correlation between the two seemingly independent physical events through the ionization parameter ($\mu$).

We have plotted the difference of breaking time between the onsets of decay process of the two events ($\Delta t$) (figure 7 - 10) as well as their higher and lower order differences ($\left[T_b^v(\mu)\right]^p - \left[T_b^s(\mu)\right]^p$, where $p = 0.1, 0.2, \ldots, 4$.) and found that the following single Fourier series represents the existence of a unique relationship, for all values of p.

$$\left[T_b^v(\mu)\right]^p - \left[T_b^s(\mu)\right]^p = f_p(\mu) = a_0 + a_1\cos(\mu w) + b_1\sin(\mu w) + a_2\cos(2\mu w) + b_2\sin(2\mu w) + a_3\cos(3\mu w) + b_3\sin(3\mu w) + a_4\cos(4\mu w) + b_4\sin(4\mu w) + a_5\cos(5\mu w) + b_5\sin(5\mu w) + a_6\cos(6\mu w) + b_6\sin(6\mu w)$$

Table 1: The coefficients with $2\sigma$ confidence bound are as follows for $< 1$:

|       | $f_{0.1}(\mu)$ | $f_{0.2}(\mu)$ | $f_{0.3}(\mu)$ | $f_{0.4}(\mu)$ | $f_{0.5}(\mu)$ | $f_{0.6}(\mu)$ | $f_{0.7}(\mu)$ | $f_{0.8}(\mu)$ | $f_{0.9}(\mu)$ |
|-------|---------|---------|----------|----------|----------|----------|----------|----------|----------|
| $a_0$ | -0.0153 | -0.0259 | -0.03236 | -0.03547 | -0.03609 | -0.03494 | -0.03267 | -0.02974 | -0.02652 |
| $a_1$ | 0.01205 | 0.01559 | 0.01486  | 0.01231  | 0.009261 | 0.006384 | 0.003963 | 0.002075 | 0.0006941 |
| $a_2$ | -0.0003276 | 0.001138 | 0.003037 | 0.004719 | 0.005933 | 0.006638 | 0.006896 | 0.006801 | 0.006456 |
| $a_3$ | -0.02253 | -0.03063 | -0.03104 | -0.02775 | -0.02307 | -0.01821 | -0.01379 | -0.01007 | -0.007072 |
| $a_4$ | -0.01301 | -0.01787 | -0.0183 | -0.01656 | -0.01395 | -0.01118 | -0.008614 | -0.006414 | -0.004619 |
| $a_5$ | 0.002717 | 0.004025 | 0.004484 | 0.004452 | 0.004156 | 0.003734 | 0.00327 | 0.002813 | 0.002388 |
| $a_6$ | 0.001779 | 0.002321 | 0.002232 | 0.001864 | 0.001415 | 0.0009831 | 0.0006144 | 0.0003226 | 0.0001054 |
| $b_1$ | -0.02081 | -0.02663 | -0.02499 | -0.0202 | -0.01462 | -0.009408 | -0.005076 | -0.001746 | 0.0006426 |
| $b_2$ | -0.01241 | -0.01476 | -0.01244 | -0.008439 | -0.00431 | -0.0007706 | 0.001934 | 0.003801 | 0.00494 |
| $b_3$ | -0.004345 | -0.005972 | -0.006119 | -0.005534 | -0.004653 | -0.003716 | -0.002849 | -0.002105 | -0.001498 |
| $b_4$ | 0.006595 | 0.008781 | 0.00869 | 0.007562 | 0.006088 | 0.004626 | 0.003342 | 0.002294 | 0.001483 |
| $b_5$ | 0.001271 | 0.000929 | -1.744e-05 | -0.001045 | -0.001911 | -0.002531 | -0.002901 | -0.003055 | -0.003043 |
| $b_6$ | 0.00123 | 0.001573 | 0.001472 | 0.001181 | 0.00084 | 0.0005212 | 0.0002552 | 5.032e-05 | -9.668e-05 |
| $w$ | 21.05 | 21.05 | 21.05 | 21.05 | 21.05 | 21.05 | 21.05 | 21.05 | 21.05 |

Table 2: The coefficients with $2\sigma$ confidence bound are as follows for $\geq 1$:

| | $f_1(\mu)$ | $f_{1.1}(\mu)$ | $f_{1.2}(\mu)$ | $f_{1.3}(\mu)$ | $f_{1.4}(\mu)$ | $f_{1.5}(\mu)$ | $f_2(\mu)$ | $f_3(\mu)$ | $f_4(\mu)$ |
|---|---|---|---|---|---|---|---|---|---|
| $a_0$ | 0.02948 | -0.0201 | -0.01718 | -0.01453 | -0.01219 | -0.01015 | -0.003732 | -0.0003901 | -0.0003901 |
| $a_1$ | 0.01907 | -0.0008617 | -0.001207 | -0.001366 | -0.001395 | -0.001341 | -0.0007075 | -8.325e-05 | -8.325e-05 |
| $a_2$ | -0.07409 | 0.00536 | 0.004736 | 0.00412 | 0.003537 | 0.003003 | 0.001169 | 0.0001239 | 0.0001239 |
| $a_3$ | -0.02795 | -0.003034 | -0.001783 | -0.0009054 | -0.0003124 | 7.011e-05 | 0.0004623 | 8.9e-05 | 8.9e-05 |
| $a_4$ | 0.02288 | -0.002136 | -0.001342 | -0.000771 | -0.0003724 | -0.0001042 | 0.0002489 | 5.936e-05 | 5.936e-05 |
| $a_5$ | 0.01247 | 0.001674 | 0.001387 | 0.001144 | 0.0009402 | 0.0007698 | 0.0002736 | 3.105e-05 | 3.105e-05 |
| $a_6$ | -0.005428 | -0.0001462 | -0.0002049 | -0.0002337 | -0.0002414 | -0.000235 | -0.0001338 | -1.948e-05 | -1.948e-05 |
| $b_1$ | -0.09546 | 0.003191 | 0.003674 | 0.003816 | 0.003729 | 0.003497 | 0.001815 | 0.0002456 | 0.0002456 |
| $b_2$ | -0.02969 | 0.005628 | 0.005459 | 0.005099 | 0.004631 | 0.004114 | 0.001841 | 0.000217 | 0.000217 |
| $b_3$ | 0.05613 | -0.0006592 | -0.0003927 | -0.0002024 | -7.121e-05 | 1.541e-05 | 0.0001131 | 2.545e-05 | 2.545e-05 |
| $b_4$ | 0.0217 | 0.0004516 | 0.000159 | -3.034e-05 | -0.0001444 | -0.0002055 | -0.0001754 | -2.214e-05 | -2.214e-05 |
| $b_5$ | -0.01012 | -0.0027 | -0.002446 | -0.002174 | -0.001901 | -0.001641 | -0.0006826 | -7.975e-05 | -7.975e-05 |
| $b_6$ | -0.008743 | -0.000252 | -0.0002798 | -0.0002861 | -0.0002776 | -0.0002598 | -0.0001375 | -2.007e-05 | -2.007e-05 |
| $w$ | 15.79 | 21.05 | 21.05 | 21.05 | 21.05 | 21.05 | 21.05 | 21.05 | 21.05 |

The profile holds the key to establish how the two independently occurring events viz. dust void and dust soliton can be linked up through plasma ionization parameter $\mu$.

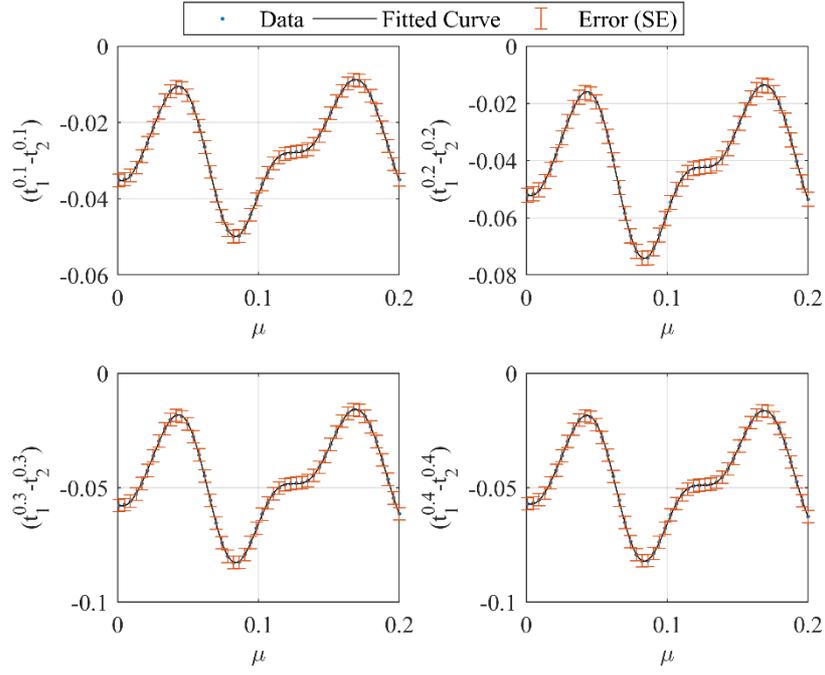

FIG. 7. The lower order ($p < 1$) correlations ranging $p$ = 0.1 to 0.4 with a maximum error of ±5% of the largest $\Delta t$ value assuming constant probability for the error within the limits of the standard error ($\sigma_\mu = \sigma/\sqrt{n}$, where $\sigma$ = standard deviation and $n$ is the total number of data points)

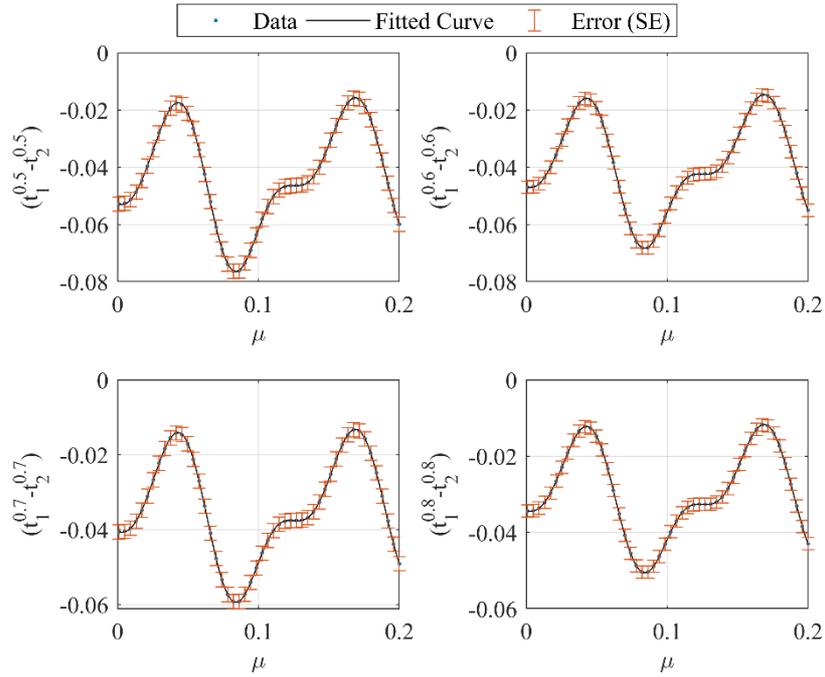

FIG. 8. The lower order ($p < 1$) correlations ranging $p$ = 0.5 to 0.8 with a maximum error of ±5% of the largest $\Delta t$ value assuming constant probability for the error within the limits of the standard error ($\sigma_\mu = \sigma/\sqrt{n}$, where $\sigma$ = standard deviation and $n$ is the total number of data points)

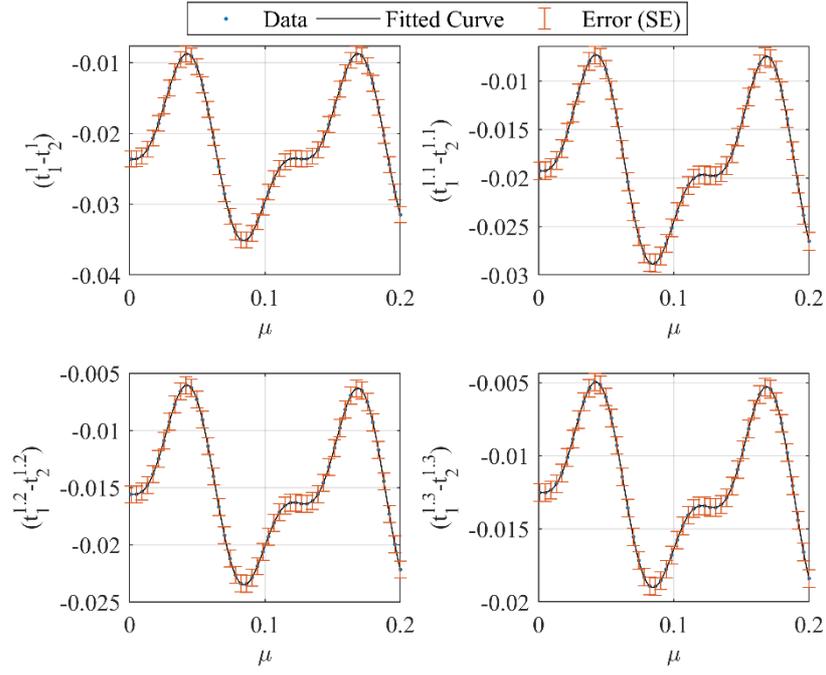

FIG. 9. The higher order ($p > 1$) correlations ranging $p = 1$ to $1.3$ with a maximum error of $\pm 5\%$ of the largest $\Delta t$ value assuming constant probability for the error within the limits of the standard error ($\sigma_\mu = \sigma / \sqrt{n}$, where $\sigma$ = standard deviation and $n$ is the total number of data points)

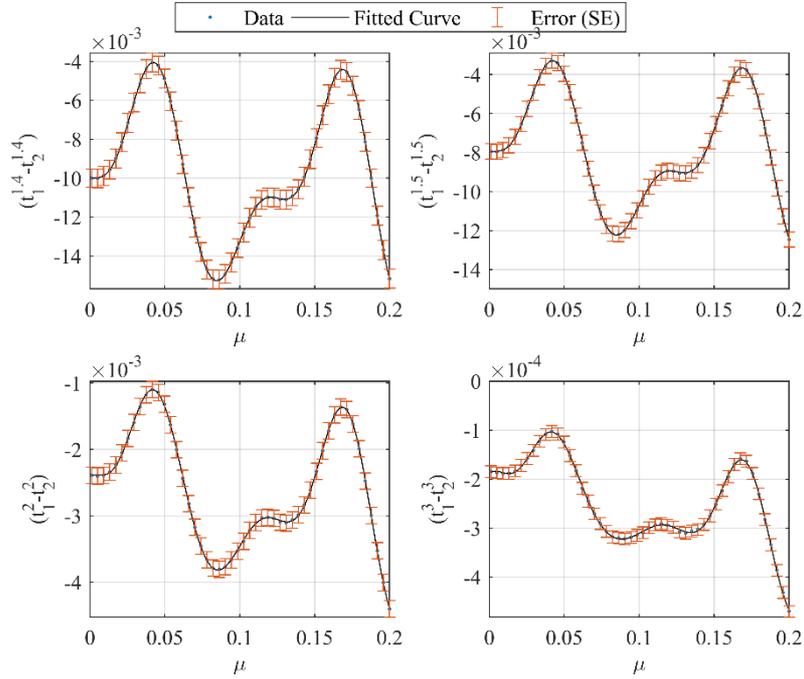

FIG. 10. The higher order ($p > 1$) correlations ranging $p = 1.4$ to $3$ with a maximum error of $\pm 5\%$ of the largest $\Delta t$ value assuming constant probability for the error within the limits of the standard error ($\sigma_\mu = \sigma / \sqrt{n}$, where $\sigma$ = standard deviation and $n$ is the total number of data points)

Since the phenomena described here are supposed to be intertwined, we further tried to find out if any such correlation exists for the coefficients also. Surprisingly the coefficients do have a correlation (figure 11 - 12) as follows,

$$g(\mu) = a'\exp(b'\mu) + c'\exp(d'\mu)$$

the coefficients with $2\sigma$ confidence bound are as follows for $p < 1$:

|     | $a_0$    | $a_1$     | $b_1$    |
|-----|----------|-----------|----------|
| $a'$ | 4.552    | -0.3169   | 0.7917   |
| $b'$ | -2.146   | -5.317    | -5.62    |
| $c'$ | -4.551   | 0.3166    | -0.7909  |
| $d'$ | -2.101   | -4.645    | -5.119   |

the coefficients with $2\sigma$ confidence bound are as follows for $p > 1$:

|     | $a_0$      | $a_1$       | $b_1$       |
|-----|------------|-------------|-------------|
| $a'$ | -0.1766    | -0.003579   | 0.008251    |
| $b'$ | -1.914     | -0.7401     | -0.5232     |
| $c'$ | 2.562e+13  | 1.556e+12   | -2.52e+13   |
| $d'$ | -33.77     | -31.95      | -33.16      |

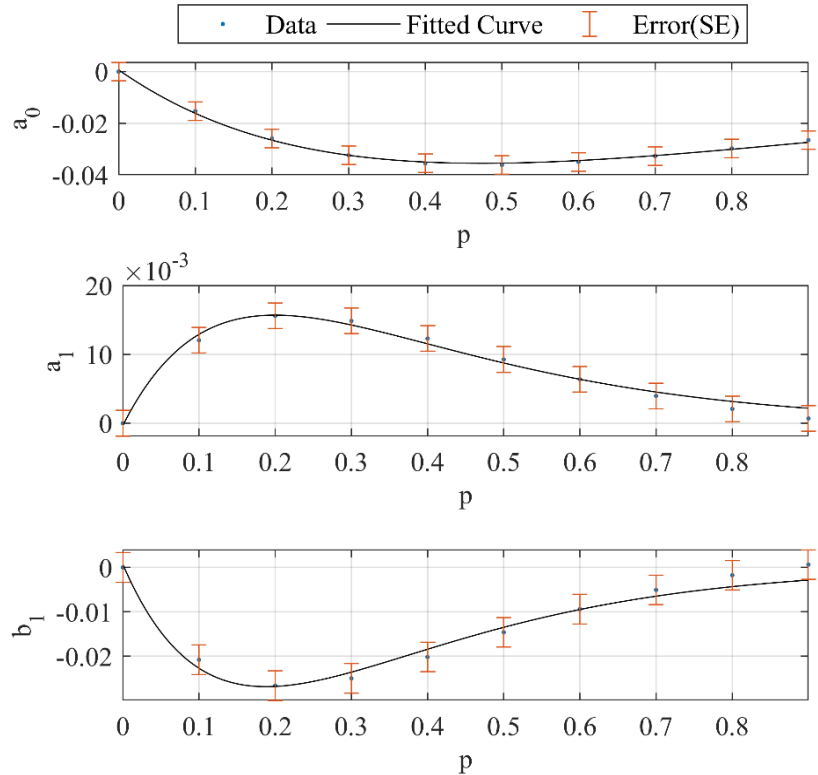

FIG. 11. The correlation between the first and second order coefficients ($a_0$, $a_1$, and $b_1$) for $p < 1$ with a maximum error of $\pm 5\%$ of the largest $\Delta t$ value assuming constant probability for the error within the limits of the standard error ($\sigma_\mu = \sigma/\sqrt{n}$, where $\sigma$ = standard deviation and $n$ is the total number of data points)

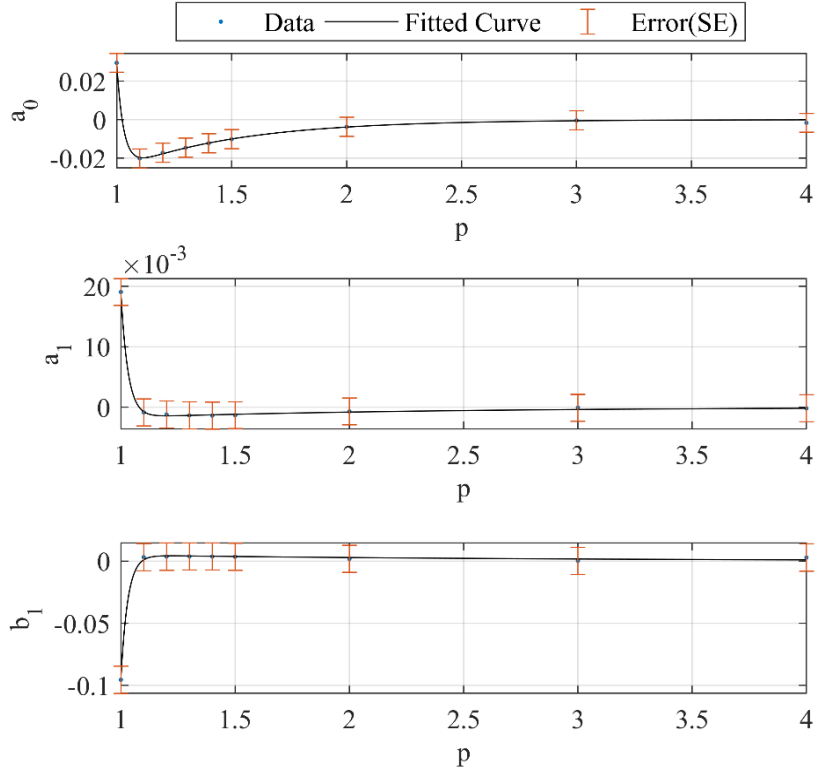

FIG. 12. The correlation between the first and second order coefficients ($a_0$, $a_1$, and $b_1$) for $p > 1$ with a maximum error of $\pm 5\%$ of the largest $\Delta t$ value assuming constant probability for the error within the limits of the standard error ($\sigma_\mu = \sigma/\sqrt{n}$, where $\sigma$ = standard deviation and $n$ is the total number of data points)

The analysis (Table 1 & 2) shows that for $p < 1$ and $p \cong 1$ the coefficients vary significantly while for higher values of $p$ ( $= 3$ or $4$) the fit coefficients remain almost same reminding the phenomena of multiscaling observed in high-resolution spectral simulation of stochastically forced Burgers equation[18] and in fluid [19]and MHD turbulence[20].

The primary cause of such multiscaling effect is probably due to shocks that get generated during the breaking of soliton or void and this indicates similarity with fluid systems explained through Navier-Stokes equation[21].

In order to verify that only the ionization parameter $\mu$ affects the correlation between the two events, we have carried out similar studies on other parameters namely, the temperature ratio ($T_d/T_e$, $T_i/T_e$), dust-neutral collision frequency($\nu_{dn}$) etc. and observed that these parameters do not have any noticeable effect. Hence, these results from the study have not been included to make the paper incisive. The present study however, without any ambiguity has established that the ionization parameter strongly affects both the phenomena individually and distinctly.

## V. SUMMARY AND DISCUSSION

An unmagnetized homogenous dusty plasma model has been considered to study two nonlinear events viz. dust void and soliton simultaneously in the same system. The time-dependent model depicts the evolution of dust density profile in the form of dust void and dust soliton when investigated separately. In case of a dust void, it is observed that the electric field produced due to ionization expels dust particles completely from the central region. Secondly, one can calculate the dust void boundary (void size) from those points where electrostatic force balances the ion drag force. The model depicts that the balance between the electrostatic and ion drag force gets disturbed for the higher value of $\mu$ and this leads to destabilization of dust void. The KdV equation derived for this problem has been used to study the existence of solitary wave structure during the evolution of dust void and it has been shown that the solitary wave structure also depends on the ionization parameter ($\mu$).

## VI. CONCLUSIONS

In physics, two independent physical events are generally correlated using statistical methods by identifying a crucial parameter for which both the events exhibit some commonality. The work presented here has shown for the first time that the plasma ionization parameter $\mu$ is capable to correlate the decay process of two independent nonlinear events viz. the dust void and dust solitons in a homogeneous dusty plasma system. Here, the time at which soliton starts to decay acts as a precursor to the onset time for a dust void decay and therefore this time lag is important to pronounce the destabilization of a dust void. The relationship between the time lag in the decay process of a soliton and dust void with ionization parameter is found to be represented by a Fourier series. Such an outcome is quite surprising since the physical mechanism for the evolution of dust void and dust soliton in a given complex plasma system are apparently independent. The present paper therefore provides a mechanism to link the instability of two seemingly different nonlinear processes and paves the way to identify otherwise independent nonlinear phenomena occurring in nature which could be correlated on the basis of this model.